\documentclass[twoside,a4paper,11pt]{sea10}
\pdfoutput=1
\usepackage{graphicx}
\usepackage{hyperref}
\usepackage{movie15}
\begin{document}
\pagenumbering{arabic}
\pagestyle{myheadings}
\thispagestyle{empty}
{\flushleft\includegraphics[width=\textwidth,bb=58 650 590 680]{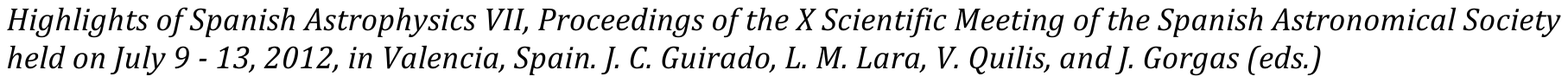}}
\vspace*{0.2cm}
\begin{flushleft}
{\bf {\LARGE
%
%%% TITLE of the paper. 
%%% TITLE of the paper. 
OSIRIS/GTC: status and prospects
%
% Do not delete next few lines
}\\
\vspace*{1cm}
%
%%% Include here the LIST OF AUTHORS.
%%% Include here the LIST OF AUTHORS.
%%% Note that the last author has to be preceeded by an AND.
J. Cepa$^{1,2}$,
\'A. Bongiovanni$^{1,2}$,
A.M. P\'erez Garc\'{i}a$^{1,2}$,
A. Ederoclite$^{3}$,
J.I Gonz\'alez-Serrano$^{4}$,
J.J. Gonz\'alez$^{5}$,
M. S\'anchez-Portal$^{6}$,
E.J. Alfaro$^{7}$,
and
A. Cabrera-Lavers$^{8}$
%
% Do not delete next few lines
}\\
\vspace*{0.5cm}
%
%%% AFFILIATIONS LIST.
%%% and the AFFILIATIONS LIST. Note that one affiliation per line.
%%% Add as many affiliations as necessary. 
$^{1}$
Instituto de Astrof\'{i}sica de Canarias (IAC), E-38200 La Laguna, Tenerife, Spain\\
$^{2}$
Departamento de Astrof\'{i}sica, Universidad de La Laguna (ULL), E-38205 La Laguna, Tenerife, Spain\\
$^{3}$
Centro de Estudios de F\'{i}sica del Cosmos de Arag\'on, E-44071 Teruel, Spain\\
$^{4}$
Instituto de F\'{i}sica de Cantabria, CSIC-Universidad de Cantabria, E-39005 Santander, Spain\\
$^{5}$
Instituto de Astronom\'{i}a, Universidad Nacional Aut\'onoma de M\'exico, A.P. 70-264, Cd. Universitaria,
04510, M\'exico\\
$^{6}$
Herschel Science Center, ESAC/INSA, P.O. Box 78, E-28691 Villanueva de la Ca\~nada, Spain\\
$^{7}$
Instituto de Astrof\'{i}sica de Andaluc\'{i}a (CSIC), Apdo. 3004, E-18080 Granada, Spain\\
$^{8}$
Grantecan S.A., E-38712 Bre\~na Baja, Santa Cruz de Tenerife, Spain\\

%
% Do not delete next few lines
\end{flushleft}
%
% Headings
\markboth{
%%% Type the SHORT version of the paper title.
%%% Type the SHORT version of the paper title.
OSIRIS/GTC: status \& prospects
}{ % Do not delete
%
%%%  First Author \& Second Author   OR   First-author et al. 
%%%  First Author \& Second Author   OR   First-author et al. if the author list 
%%% contains three or more authors.
Cepa et al.
% 
% Do not delete next few lines
}
\thispagestyle{empty}
\vspace*{0.4cm}
\begin{minipage}[l]{0.09\textwidth}
\ 
\end{minipage}
\begin{minipage}[r]{0.9\textwidth}
\vspace{1cm}
\section*{Abstract}{\small
%
% ABSTRACT ABSTRACT ABSTRACT
% ABSTRACT ABSTRACT ABSTRACT
%%% Type the ABSTRACT of your paper
OSIRIS is the optical Day One instrument, and so far the only Spanish instrument, currently
operating at the GTC. Building and testing an instrument for a 8-10m-class telescope with nonprevious
commissioning in turn, has represented a truly unique experience. In this contribution,
the current status, the last commissioning results and some future prospects are given.
%
% Do not delete next few lines
\normalsize}
\end{minipage}
%
%
%%% BODY of the paper
%%% BODY of the paper
% 
\section{Introduction \label{intro}}

OSIRIS is the imaging system (with either broad- and narrow-band via tunable filters) and a low-resolution 
long-slit \& multi-object spectrograph operating since the Day One at the Nasmyth ``A'' focus of the GTC. 
The instrument works in the wavelength range from 365 to 1000 nm with an unvignetted FOV of 
$7.8\times 7.9$ and $7.8\times 5.2$ arcmin$^2$ in direct imaging and spectroscopy, respectively 
(\cite{cepa03}). 

The Scientific Commissioning of OSIRIS begun in December 2008, almost simultaneously to the one of the GTC.
Since such date until this time, several observing modes of the instrument has been commssioned, that is, 
broad-band imaging, narrow-band imaging with the Red (651-935 nm) Tunable Filter (RTF) and long-slit spectroscopy. 
Although they are not still fully commissioned, the fast photometry observing mode, multi-object 
spectroscopy (MOS) and the Blue (375-675 nm) Tunable Filter (BTF) have been tested. Remaining -transversal-
possibilities (Nod+Shuffle, $\mu$-shuffling and Multiplex) are subject to the completion of
the MOS commissioning. This contribution contains the highlights of the instrument status (hardware
\& software) and preliminary results of the observing modes in commssioning phase. 

\section{OSIRIS  - Hardware} 

Essentially, the OSIRIS main hardware in the focal station at the telescope is fully operational. The Mask Magazine was repaired, and is currently operating with the maximum load of 13 masks. However, some spurious faults of the RTF and electromagnetic interferences between controllers of the Red and Blue TFs have been detected. It has been decided to operate with only one of the TF switched--on at a time for avoiding interferences. However, switching--off one TF and on the other take only few minutes, and the calibration has found to be stable even after switching a TF off and back on. 

On the other hand, the technical commssioning of the Mask Driller machine is under way at the supplier headquarters and the acceptance by OSIRIS team is scheduled for December 2012. New linear encoders for higher accuracy have been installed and tests are under way.

Finally, the purchase of the remaining Order-Sorting filters (365 to 450 nm) for the BTF is being handled, and the final manufacturer has been selected. 
 
\section{OSIRIS - Software} 

The appropriate operation of the instrument, as well as the common procedures for the reduction of OSIRIS data depends on different software applications whose final develoment and tuning are still on going or near completion. 

The first example is the RTF Calculator (as a part of the
OSIRIS SNR Calculators and TF Setup tool system, developed by J.I. Gonz\'alez-Serrano). The calculators have demonstrated high reliability since the beginning of the scientific operations. A new calibration has been implemented in the TF Setup tool, according to new calibration (Gonz\'alez et al., in preparation), but it is not implemented yet. This new calibration allow wavelength and FWHM tuning accuracies of 0.2 and 0.3 ${\rm \AA}$, respectively, in the whole spectral range of the RTF. The BTF calculator is being prepared.

Another software package developed during the last years is the Mask Designer (MD). This is the main tool for design and preparation of the manufacturing process of the slitmasks to be used in multi--object spectroscopy with OSIRIS. The MD is a {\tt Java} plug-in to the well-known {\tt JSky} application, described in \cite{ignacio04}, whose user's interface was finally developed with the 
{\tt PORIS} toolkit \cite{txinto10}. The beta version of the MD was released in 2011 and it is now undergoing different tests. The MD is an interactive application which allows a visualization of the mask design in three simultaneous domains: the mask model, the CCD mosaic, and the sky. This is possible through a same number of {\tt JSky} based editors. The user can interact with 
each editor if desired. 
%The slits can be created in any of these domains via cursor, which are 
%continually updated thanks to the polynomial transformations that link them. This approach can 
%be useful for designing engineering or special slitmasks. 
Mask designing is a user's tasks that can use either a list of equatorial coordinates with proper motions (if apply), or through a list of rectangular coordinates referred to the system of a pre-image from OSIRIS.
%This target list includes also the slit type, width, length and the preliminary assigned priority. Each
%telescope pointing in MOS involves one MD project, in which lives one o more masks.
As can be inferred, an accurate coordinate mapping by the MD is essential. Different coordinate transformations (via high--order polynomials) are needed to determine the position of an object in the mask from its sky position, and from the mask to the detector mosaic. In the current version of the MD, the former
transformation was derived from the nominal positions of 110 pinholes in a special mask manufactured for this purpose, while the latter comes -provisionally- from the analysis of archive OSIRIS images of
the VOph8142 field, which was astrometrically calibrated by using the USNO-B1 catalogue (\cite{monet03}).
%Three different slits can be defined in a mask: (a) circular, (b) rectangular, and (c) arc-shaped slits. 
%Fiducial stars for acquisition are normally represented in the mask by circular slits. The MD checks permanently
%for potential slits collision and mask integrity by changing the default priority of the involved slits. 
%A screenshot of the MD user's interface in the designing of a slitmask is shown in Figure~\ref{fig1}.

%\begin{figure}
%\center
%\includegraphics[scale=0.30]{cepaj_osirisF0.pdf}
%\caption{\label{fig1} Screenshot of the Mask Designer Tool - Main Panel in the Manufacturing fashion. 
%The background shows a fraction of the Sky Editor with some designed slits over selected targets. The 
%shady regions in the background correspond to the envisaged imprint of the model 2D spectra.
%}
%\end{figure}
 
Last but not least, the OSIRIS Offline Pipeline Software (OOPS, \cite{ederoclite11}) is the package 
whose development has held the team's efforts since the beginning of the OSIRIS Commissioning. 
The Pipeline is based on
{\tt Python} scripts with nested {\tt Pyraf} and {\tt NumPy} tasks, and it is oriented to solve
automatically the main common procedures in the reduction of OSIRIS data from its different observing 
modes. The OOPS (including the MOS mode) is in validation 
phase by users of the Extended Scientific Team, and a few minor bugs have been detected and corrected. 
Some features are not yet implemented (e.g. 1D spectra extraction), and the official scope for a
desired pipeline has been revised by the GTC Users Committee in late June 2012. On the other hand,
the stability of the OOPS depends, among other, on stable {\tt fits} headers in the output of the instrument,
a final definition of calibration lamps, and on the settling of the instrument control parameters.
All these guarantees should be provided by the proprietary of the instrument. Is expected that the public OOPS release will be available via OSIRIS blog 
(\href{http://gtc-osiris.blogspot.com.es/}{here}) in the last months of 2012.
 
\section{MOS: the first tests}

Since November 2011, a preliminary version of an automated script for MOS target acquisition at the GTC is 
operating. This application calculates the offsets and rotation needed to align the positions of
the fiducial marks, in the detector mosaic domain, with the corresponding positions of acquisition stars. 
The release of the beta version of the MD met by chance with the early tests of the automated 
acquisition procedure at the GTC. Thus, it was decided to design four test masks using 
both equatorial coordinates lists (two masks) and pixel positions from OSIRIS pre--images, in order to obtain 
useful information for the improvement of the designing, manufacturing and acquisition processes. 
%As can be 
%inferred, an accurate coordinate mapping by the MD is essential. Different coordinate transformations
%(via high-order polynomial) are needed to carry out the position of an object on the sky in the
%mask, and from the mask to the detector mosaic. In the current version of the MD, the former
%transformation was derived from the nominal positions of 110 pinholes in a special mask manufactured
%for this purpose, while the latter comes -provisionally- from the analysis of archive OSIRIS images of 
%the VOph8142 field, which was astrometrically calibrated by using the USNO-B1 catalogue (\cite{monet03}).

\begin{figure}[h]
\begin{center}$
\begin{array}{cc}
\includegraphics[scale=0.31]{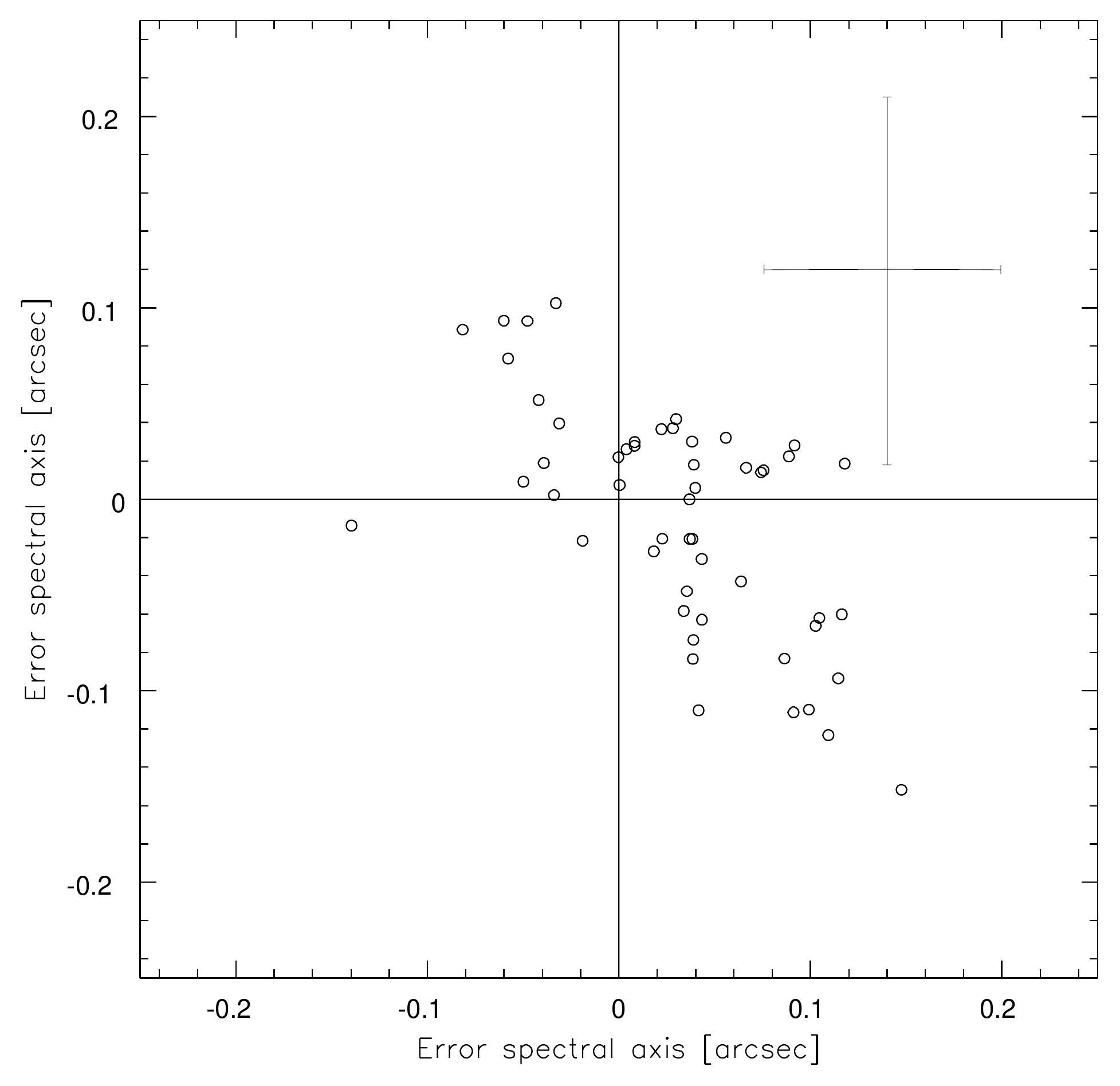} &
\includegraphics[scale=0.31]{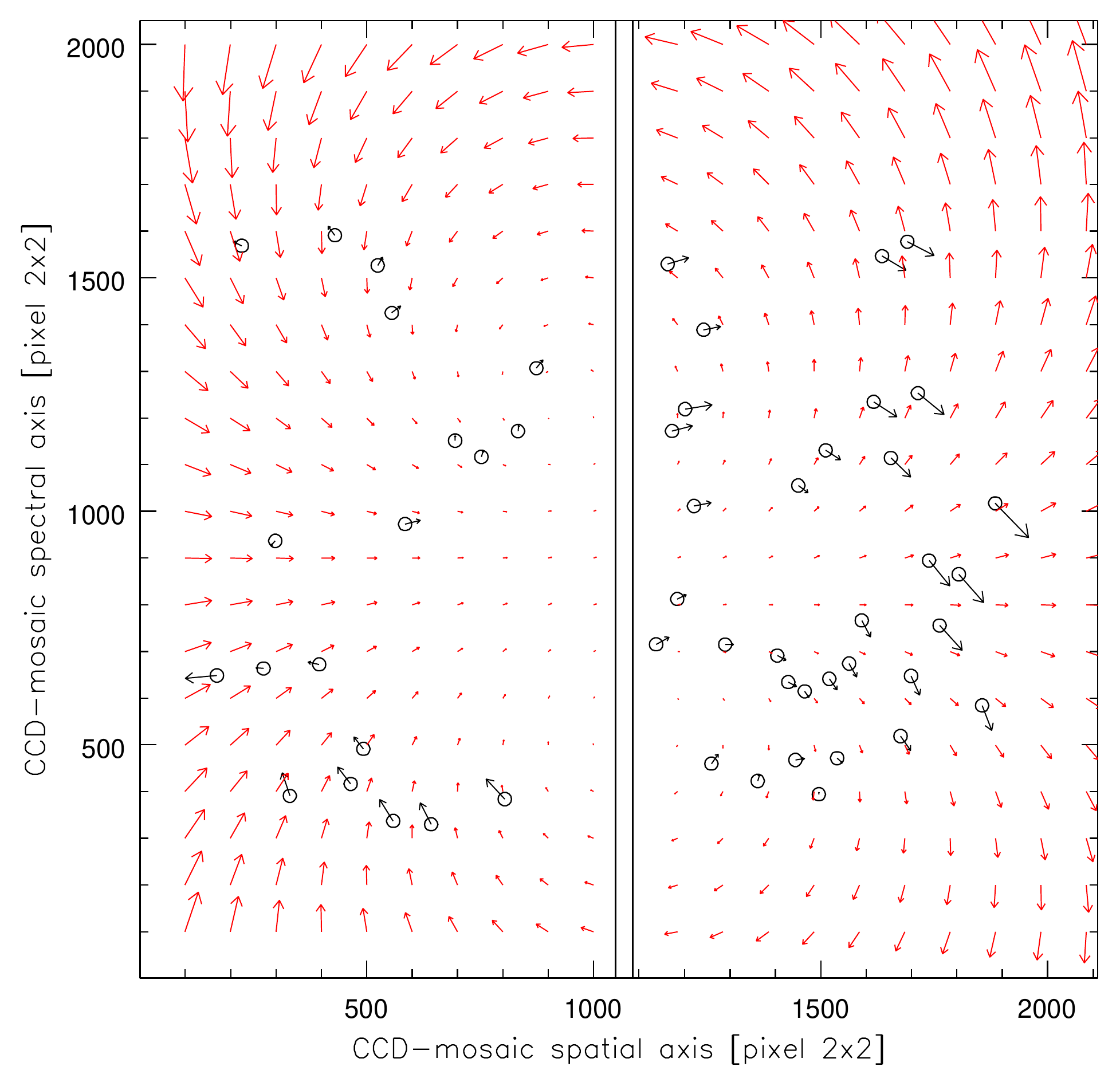} 
\end{array}$
\end{center} 
\caption{\label{fig1} Positional errors distribution (left), obtained by comparing the projected positions
on the mosaic and the measured positions of the slits in a flat frame. The bars in the first quadrant represent
the maximum positional deviation of the slits after manufacturing. The map at the right side represents
such errors magnified 400 times, together with the OSIRIS known geometrical distortions (red arrows) magnified
by a factor 5. It is clear that there are not a dependence of the errors on such distortions.
}
\end{figure}

%All masks were manufactured with a CNC milling machine at the IAC workshop in a mean period of 14
%days each one. The controlled measurement of the finished masks revealed errors (scaled to the sky) 
%up to 0.1 arcsec. It is expected that the operations of a dedicated milling machine will start soon, 
%so that these positional deviations and the manufacturing period would decrease significantly.

The equatorial coordinates of 53 stars (V$<$20, from \cite{yadav08}) in a selected field of the Galactic 
cluster M67 were selected to build the input list for the first test slitmask, whereas the
the following test masks were prepared using either OSIRIS pre--images on the
field of two galaxy clusters (Abell 2219 and CL0024+17), or a target list given in celestial coordinates. 
In such cases, the target galaxies were selected with R$<$22.5.
From February to August 2012 the slitmasks were tested at the instrument. Unfortunately, bad seeing 
conditions dominate during the acquisition of the M67 field, while in the case of the three remaining 
masks there arose the need to reduce the sky background level using broad-band filters when 
the pre-images were obtained, and also when the corresponding acquisition was accomplished, since due to
scheduling conflicts at the telescope, the tests necessarily had to be performed in bright time.
%The distortions of the broad-band filters have not yet been studied in detail, so it was not possible 
%to correct the measured positional errors of the slits in the last three slitmasks.
Even so, some interesting results could be derived from the tests, from which the following are the more remarcable ones:

\begin{itemize}
 
\item{After a first--order correction for acquisition residual ($\leq$0.15 arcsec), the maximal amplitude of the
positional errors of the slits in the first mask, obtained by comparing the projected positions on the mosaic
and the measured positions of the slits in a flat frame, was $\pm$0.1 arcsec. Figure~\ref{fig1} (left) depicts 
the distribution of these errors, which are also represented at the right side of the same Figure but in
a map of the detector mosaic, together with the geometrical distortions measured in the whole OSIRIS FOV. 
Such errors are a prediction of those that could be obtained in the case of a mask with a design based on 
a pre--image, and they are independent of higher-order effects.
}
  
\item{By comparing the position of the targets in the acquisition image with that of the slits in a flat 
frame, the positional errors are encircled by an envelope $\sim$2 times bigger than the maximum
error in Figure~\ref{fig1} (left). Taking into account the manufacturing errors, as well as the thermal fluctuation, 
mechanical flexures, and the positional errors of the targets in the input list, the final position of 
the slits in the mask must be better than 6\% of the slit width \cite{ignacio04}. Thus, for the slits 
in the first test mask (width=1 arcsec) the maximum measured amplitude of the positional deviations should be
0.1 arcsec, that is, the maximum positional error at present is slightly more than $\sim$0.1 arcsec above specs. 
}
 
\item{Because the limitations pointed out above, the tests with masks based on pre--images are inconclusive in what refers to the estimation of positional errors.}  

\end{itemize}
 
\section{Some examples of OSIRIS exploitation}

\begin{figure}[h]
\begin{center}$
\begin{array}{cc}
\includegraphics[scale=0.36]{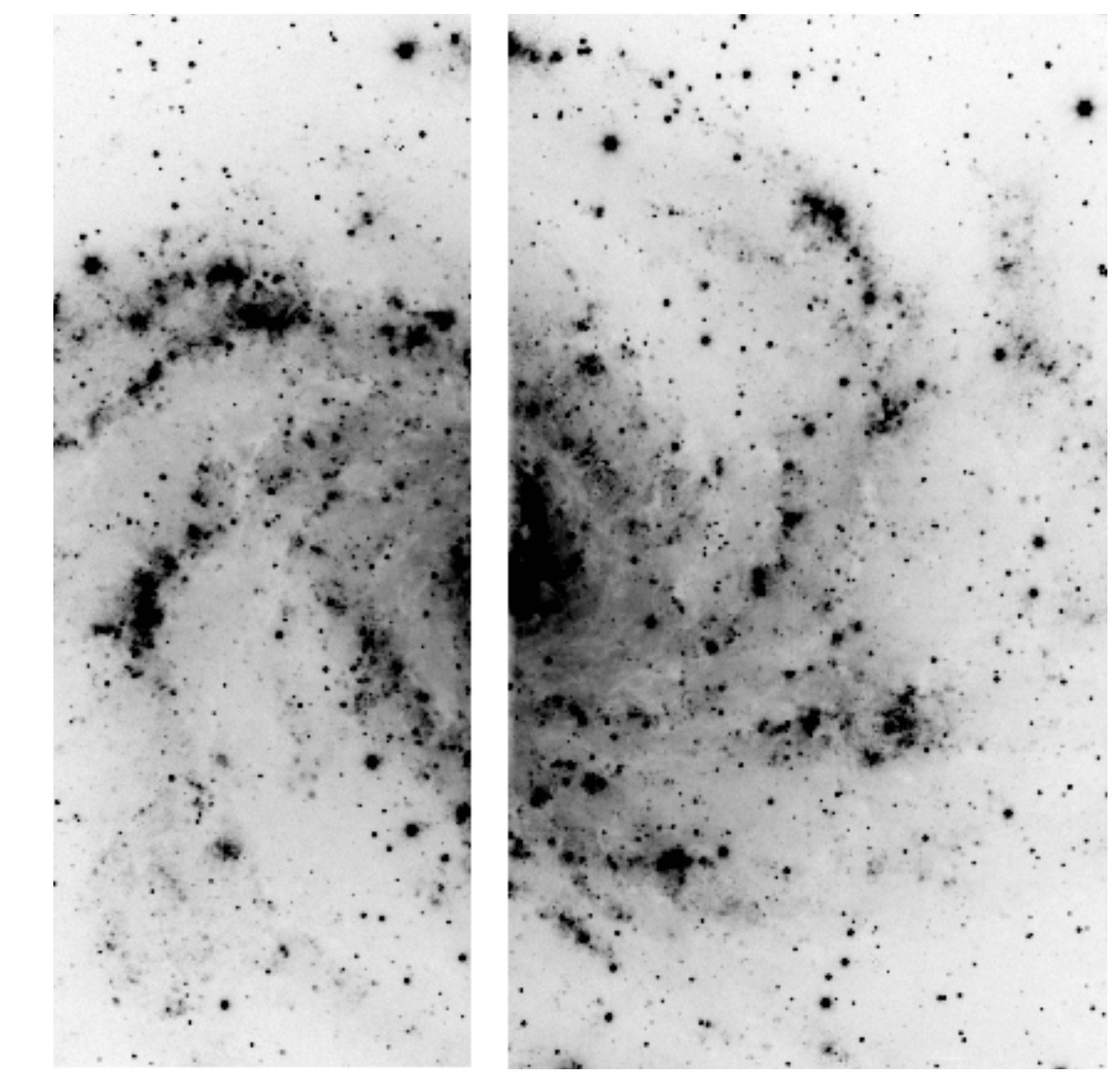} &
\includegraphics[scale=0.36]{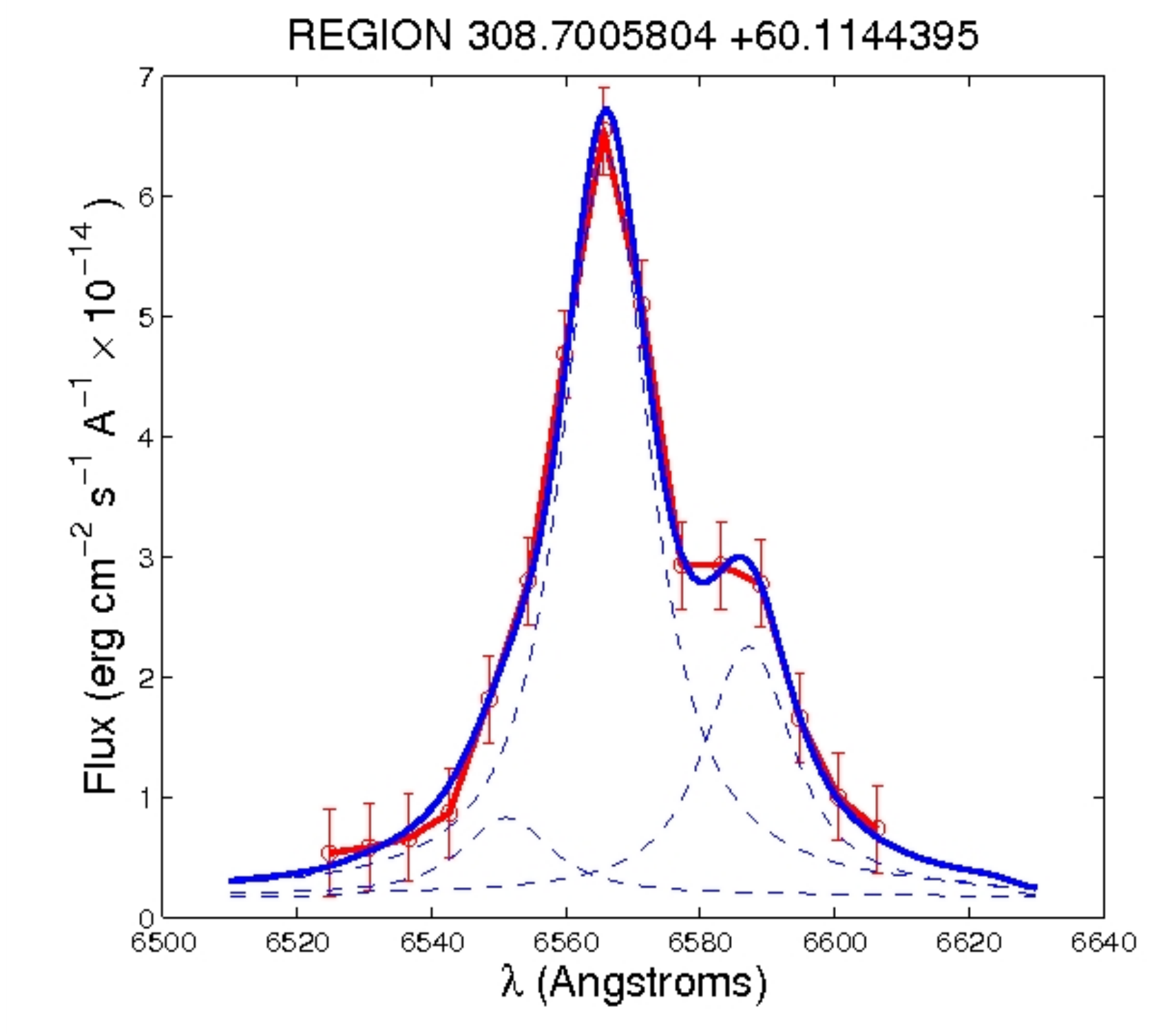} 
\end{array}$
\end{center}
\caption{\label{fig2} Left: Coadded image from an H$\alpha$+[NII] RTF scan of NGC 6946. Right: Pseudo--spectrum of a selected HII region of NGC 6946 (red), the sum of three Voigt profiles fitted to each emission line (blue), and the Gaussian profiles resulting from its deconvolution of the Airy profile (dashed line). Courtesy of B. Cedr\'es.
}
\end{figure}

The OSIRIS Tunable Filters provides GTC with extraordinary capabilities for low-resolution 2D
spectroscopy. One example is the OTELO project (see the related contribution in this              
volume), which is now furnishing the flux-deepest emission line object survey of the universe
up to z$\sim$7. Another example of the OSIRIS scientific advantages is the 2D spectroscopy
of extended sources, i.e., the HII regions of the nearby spiral galaxy NGC 6946 (Cedr\'es et al.
in preparation). Figure~\ref{fig2} (left) shows the synthesis of an H$\alpha$+[NII] RTF scan 
of this galaxy. The successful strategy of investing observing time in the outfield sampling 
of airglow emission lines, allowed a very accurate subtraction of these features. Subsequent
analysis of the scan slices provided pseudo-spectra of the isolated HII regions. The 
Figure~\ref{fig2} (right) depicts the observed pseudo-spectrum of a selected HII region in NGC 6946 
and its deconvolution of the instrumental profile.  

Even with the position errors detected in the MOS tests, the masks designed from a pre--image
of the galaxy cluster fields were useful for obtaining scientific quality data. The slit width for
the targets was fixed in 1.2 arcsec, and the reduced spectra were obtained from the coaddition of
2$\times$1200s integrations with the OSIRIS R1000R grism. The raw 2D spectra of the selected galaxies in
the field of Abell 2219 can be apreciated in Figure~\ref{fig2} (left). A sample of reduced spectra 
of this test are shown in Figure~\ref{fig2} (right).

\begin{figure}[h] 
\begin{center}$
\begin{array}{cc}
\includegraphics[scale=0.33]{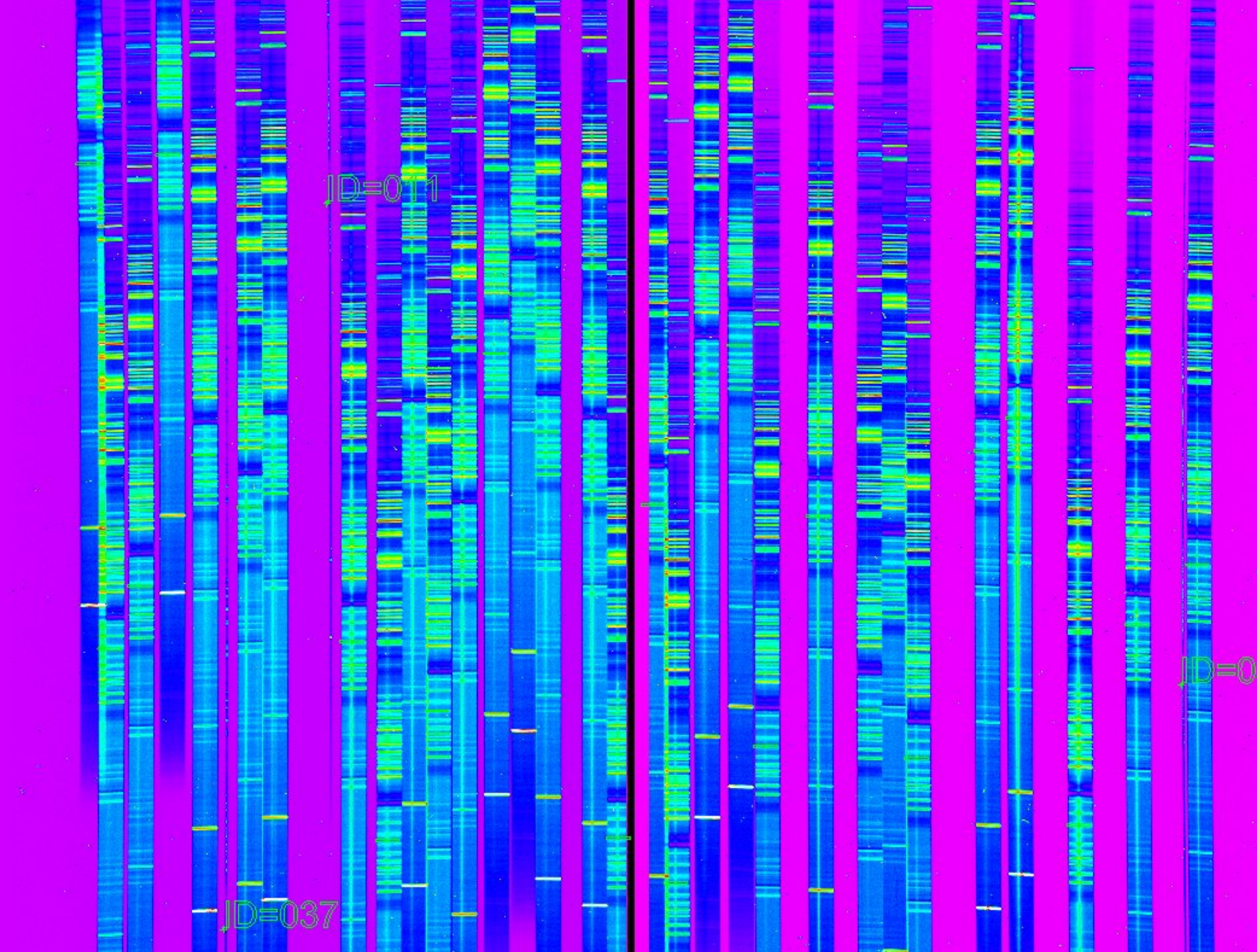} &
\includegraphics[scale=0.27]{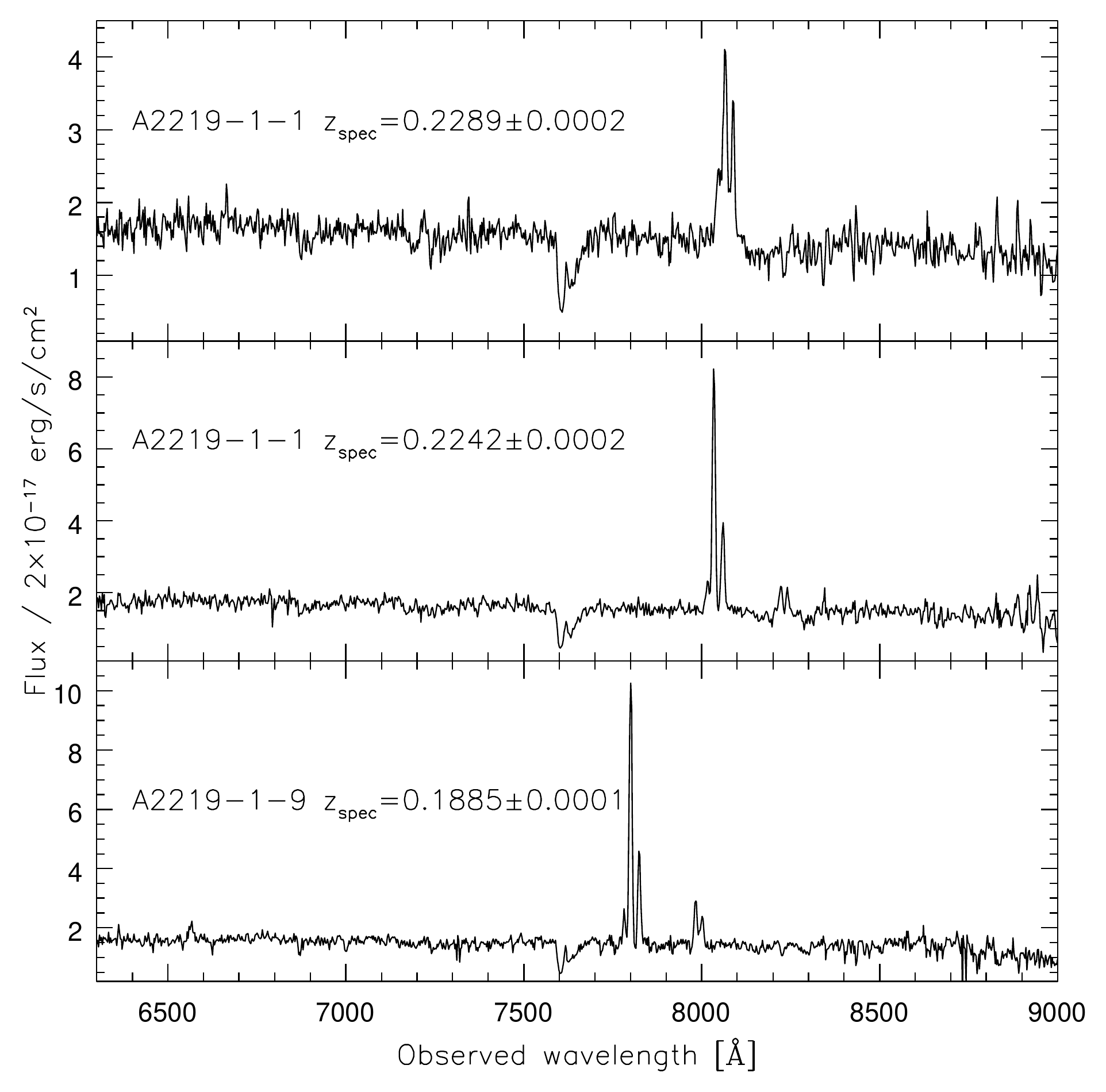} 
\end{array}$
\end{center}
\caption{\label{fig3} Raw image of a MOS test on the field of the cluster Abell 2219 (left). Some reduced OSIRIS--MOS 
spectra of selected emission line galaxies in this field (right).
}
\end{figure}
 
\section{Summary}

About the OSIRIS hardware, main pending issues are related with the remaining filters for the BTF and the delivery of the Mask Driller machine. On the other hand, a beta version of the OSIRIS Pipeline
(OOPS) is expected to be released in short. The BTF setup tool and calculator is expected to be ready
when the Commissioning of this sub-system is completed.  

Although the initial tests of the MOS have not been made under appropiate conditions, the results obtained
are so far satisfactory: it results that the MOS using pre--imaging approach could be scientifically exploited at the present time. The upgrading of the MD polynomial transformations coefficients, and thus the 
minimization of the positional errors, depends on high-quality calibration data which only can be provided by the telescope. This is one of the ongoing tasks in the MOS Commissioning.

\end{document}